\documentstyle[aps,prl,epsf]{revtex}
\tightenlines

\newcommand{\be}{\begin{equation}}
\newcommand{\ee}{\end{equation}}
\newcommand{\bea}{\begin{eqnarray}}
\newcommand{\eea}{\end{eqnarray}}

\begin{document}

\draft

\title{Hybrid Quarkonia with Dynamical Sea Quarks}

\author{CP-PACS Collaboration\\
$^1$T.~Manke, $^1$A.~Ali Khan, $^2$S.~Aoki, 
$^{1,2}$R.~Burkhalter, $^1$S.~Ejiri,
$^3$M.~Fukugita, $^4$S.~Hashimoto, $^{1,2}$N.~Ishizuka, $^{1,2}$Y.~Iwasaki, 
$^{1,2}$K.~Kanaya, $^1$T.~Kaneko, $^4$Y.~Kuramashi, $^1$K.~Nagai,
$^4$M.~Okawa, $^1$H.~P.~Shanahan,  $^{1,2}$A.~Ukawa, $^{1,2}$T.~Yoshi\'e}

\address{
$^1$Center for Computational Physics,
University of Tsukuba, Tsukuba, Ibaraki 305-8577, Japan \\
$^2$Institute of Physics, University of
Tsukuba, Tsukuba, Ibaraki 305-8571, Japan \\
$^3$Institute for Cosmic Ray Research,
University of Tokyo, Tanashi, Tokyo 188-8502, Japan \\
$^4$High Energy Accelerator Research Organization
(KEK), Tsukuba, Ibaraki 305-0801, Japan}

\date{\today}

\maketitle

\begin{abstract}
We present a dynamical lattice calculation with 2 flavours 
for bottomonium states with an additional gluonic excitation.
Using improved actions for the quarks and gauge fields at a 
lattice spacing of $a \approx 0.1$ fm,
we find 10.977(61)(62) GeV for the energy of the lowest lying 
$b\bar bg$-hybrid, where the first error is statistical and 
the second denotes the systematic uncertainty due to the determination 
of scale. In a parallel quenched simulation we demonstrate 
explicitly that vacuum polarisation 
effects are less than 10$\%$ of the splitting with the ground state.
\end{abstract}



The intense experimental search for particles with exotic quantum 
numbers has triggered theoretical interest in hadrons
which contain an excitation of the gluon flux \cite{phenom_hybrid}.
The lattice approach to QCD has already resulted 
in consistent predictions for the masses of
hybrid quarkonia from first principles 
\cite{cm_hybrid,iso_hybrid,aniso_hybrid,morning_hybrid}.
The recent results suggest that the lowest lying hybrids are 400-500 MeV
above the phenomenologically interesting $B \bar B$ threshold.
However, for numerical simplicity, all those calculations were performed
in the quenched approximation, where vacuum polarisation effects 
have been ignored. In this paper we present our results from
a full calculation in which dynamical sea quarks are included.
Dynamical results for light hybrids have been reported by the
SESAM Collaboration \cite{SESAM_light}. Here our main goal 
is to establish whether the dynamical treatment of the light sea quarks 
and the heavy valence quarks will result in large shifts of the hybrid levels.

On the lattice, hybrid mesons are particularly difficult to treat
as their excitation energies are much larger than for conventional states.
This makes it very difficult to resolve the exponential decay 
of the meson propagator on coarse lattices. It has been demonstrated
that such problems can be circumvented on lattices with a finer resolution
in the temporal direction \cite{peardon_and_morningstar}, 
but the implementation of dynamical sea quarks on anisotropic lattices 
has yet to be achieved.

This work is part of a larger project to study sea quark effects
in QCD on isotropic lattices \cite{cppacs_light}. 
For the gluon sector we chose a renormalization-group improved
action, which is written in terms of standard
plaquettes, ${\rm Tr} P_{\mu \nu}$, and $1\times 2$ rectangles, 
${\rm Tr} R_{\mu \nu}$:
\be
S_{\rm g} =  \frac{1}{g^2}\sum_{\mu\nu}
\left\{c_0 {\rm Tr} P_{\mu \nu} + c_1 {\rm Tr} R_{\mu \nu}\right\}~~~~(c_0 + 8 c_1 = 1)~~.
\label{eq:iwasaki}
\ee
Our prescription $c_1=-0.331$ is motivated by an RG-analysis of the
pure gauge theory \cite{iwasaki}. 
For the light sea quarks in the gauge field background 
we chose a lattice formulation which removes both the doublers 
and ${\cal O}(a)$ discretisation errors:
\be
S_{\rm q} = \bar q~(\slash \hskip -7pt \Delta + m_q )~q + a~\bar q~
\Delta^2~q -c_{sw}(a)~a~\frac{ig}{4} \bar q~ \sigma_{\mu\nu}F_{\mu \nu}~q~~.
\label{eq:light_fermion}
\ee
Further details and results for the conventional light hadron spectrum
can be found in \cite{cppacs_light}.

Here we study heavy hybrid states as they are of particular 
relevance for ongoing experiments at B-meson factories.
To this end we implemented a non-relativistic approach (NRQCD)
for the heavy $b$ quarks on a fine lattice with  $a \approx 0.10$ fm.
This approach is well suited, owing to the
small velocity of the quarks within the flat hybrid potential.
Unlike calculations for the spin structure in quarkonia we expect
only small corrections for spin-averaged quantities from higher order 
relativistic terms. This assumption has already been tested explicitly 
for charmonium \cite{aniso_hybrid}.
To leading order we describe the forward evolution of the heavy
quark as a diffusive problem \cite{nrqcd}
\bea
G(t+1)= \left(1-\frac{aH_0}{2n}\right)^n U_t^{\dag}\left(1-\frac{aH_0}{2n}\right)^n G(t) ~~\nonumber \\
H_0 = -  \frac{\Delta^2}{2m_b} ~~,~~ 
\delta H = - c_7 \frac{a \Delta^4}{16n m_b^2} + c_8 \frac{a^2 \Delta^{(4)}}{24m_b} ~~.
\label{eq:evolve}
\eea
The parameter $n$ was introduced to stabilise the evolution against
high-momentum modes. We chose $n=2$ throughout this analysis.
As it is common practise, we have also included the correction terms 
$c_7$ and $c_8$ to render the evolution equation (\ref{eq:evolve}) 
accurate up to ${\cal O}(a^4)$, classically. 
Radiative corrections induce terms of ${\cal O}(\alpha a^2)$ and we 
applied a mean-field improvement technique to reduce such errors as 
first suggested by Lepage {\it et al.} \cite{viability_pt}. 
We decided to divide all gauge links by the average link in Landau gauge:  
$u_0=\langle 0|(1/3)~U_\mu |0 \rangle_{\rm L}$, which is a suitable choice 
to reduce the unphysical tadpole contributions in lattice perturbation theory.

Non-relativistic meson operators can be constructed using the standard 
gauge-invariant definitions of \cite{aniso_hybrid}. For the magnetic hybrid  
this amounts to an insertion of the colour-magnetic field, $B_i$, 
into the bilinear of two-spinors:
\be
^1H(x) = \sum_{i=1}^3~\psi^{\dag}(x)~B_i~\chi(x) ~~.
\label{eq:hybrid}
\ee
In the following we will denote as $1B$ and $2B$ the ground state and
the first excitation onto which this lattice operator can project.
We also employ fuzzed link variables and several different smearings for 
the quark fields. This results in different projections of the 
source operators onto the ground state and it is useful when extracting
higher excitation with the same quantum number.

For our study we calculated the bottomonium spectrum on
400 dynamical configurations at $(\beta,\kappa)=(2.10,0.1382)$ 
on a $24^3\times 48$ lattice. 
This corresponds to the lightest sea quark mass of our full data set
at $\beta=2.1$ and we measured 
$m_\pi/m_\rho = 0.5735(48)$ \cite{cppacs_light}.
For the comparative quenched analysis we accumulated 192 independent configurations at $\beta=2.528$. This coupling was chosen so as to match the lattice spacing of the dynamical run. In both cases we find $a \approx 0.11$ fm from 
the string tension $\sqrt{\sigma}=440$~MeV. 
Here we take the $1P-1S$ splitting to set the scale. 
As expected, such a definition results in slightly different lattice spacings
for our two data sets. 

Finite size effects are known to be small for heavy quarkonia and 
they have been explicitly checked for $1S, 1P$ and $1B$ in 
charmonium \cite{aniso_hybrid}. Since here we study the bottomonium system 
on even larger lattices ($L \approx 2.5$ fm), we do not expect any 
volume dependence for the ground state masses in our analysis.
The simulation parameters for our two data sets 
are collected in Table \ref{tab:parameters}.

The results for hadron masses are obtained from correlated 
multi-exponential fits to different smearings, $\alpha$, and timeslices, $t$: 
\be
C(\alpha,t) =   \langle M_\alpha^{\dag}(t)M_\alpha (0) \rangle = \sum_{i=2}^{n_{\rm fit}} a_i(\alpha) e^{-m_i t}~~.
\label{eq:fit}
\ee
In Table \ref{tab:results} we present the results from multi-exponential fits
($n_{\rm fit} \le 4$).

\begin{figure}
\begin{center}
\leavevmode
\hbox{\epsfxsize = 10 cm \epsffile{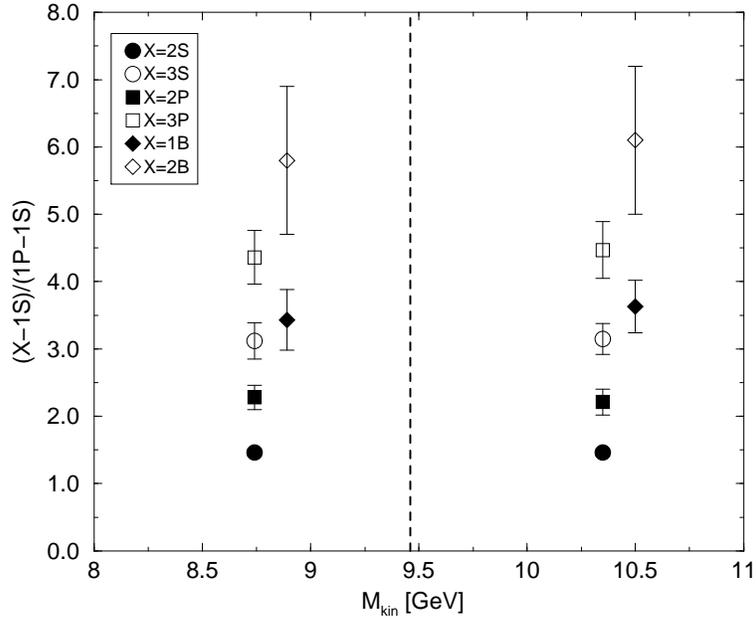}}
\end{center}
\caption{Mass independence of spin-averaged bottomonium spectrum.
Here we plot the excitations against the kinetic mass of $^{3}S_1$
from two different quark masses. The measured energies are 8.74(15) GeV and
10.35(19) GeV. The hybrid results are shifted for clarity and the
experimental Upsilon mass is shown as a vertical dashed line.}
\label{fig:RX_vs_mkin}
\end{figure}

As it is standard in lattice calculations with heavy quarkonia, we have tuned
the bare quark mass in Equation \ref{eq:evolve}, so as to reproduce the experimental
value of $M_{\rm kin}/(1P-1S) = 21.5$, where $aM_{\rm kin}$ is determined for
the non-relativistic dispersion relation of the S-state on the lattice.
In Figure \ref{fig:RX_vs_mkin} we demonstrate the mass independence of the 
spin-independent ground states at $(\beta=2.528, N_f=0)$, along with 
their higher
excitations. This independence is important as it allows us to extract
predictions at a slightly non-physical point. It also means that our result is stable against possible radiative corrections to the quark mass.
Since in NRQCD calculations all energies are measured with respect to the
ground state, $1S$, we introduced the ratio $R_X = (X-1S)/(1P-1S)$ to quote the 
normalised splitting with respect to the $1P-1S$.

Here we focus on the ratio $R_{1B} = (1B-1S)/(1P-1S)$ which determines 
the mass of the lowest lying hybrid. The velocity expansion of NRQCD works
very well for the hybrid states, owing to the slow quarks in a flat hybrid
potential. Therefore we do not expect significant changes for mass of the
spin-averaged hybrid state due to higher order relativistic corrections.
To leading order in the NRQCD Hamiltonian, the singlet state of
Equation \ref{eq:hybrid} is degenerate with the corresponding spin-triplet states 
($0^{-+},1^{-+},2^{-+}$). These are all the states with zero orbital angular
momentum and they include the exotic combination $1^{-+}$, due to
the coupling of the spin to the gluon angular momentum.
A near degeneracy was also reported for hybrid states with additional
orbital angular momentum \cite{iso_hybrid}. Such states give rise 
to more exotic states, such as $2^{+-}$.
The above-mentioned degeneracies will be lifted once higher order relativistic
corrections are re-introduced into the NRQCD Hamiltonian \cite{montpellier,ron_spin}.

The main uncertainty in lattice calculations of hybrid excitations so far 
is the absence of dynamical sea quarks. Previous estimates of quenching 
errors frequently referred to the uncertainty in the determination of 
the lattice spacing as a limiting factor of quenched simulations. However, 
it is not clear a priori, whether the high gluon content of the hybrid itself 
would cause large shifts in its mass once dynamical sea quarks are
introduced.  Comparing the results in Table \ref{tab:results}, 
we find, perhaps surprisingly, that this is not the case. 
Indeed, our quenched estimate for $R_{1B}=3.43(45)$ is in excellent 
agreement with the dynamical simulation, $R_{1B}=3.59(14)$. 
This means that quenching errors are smaller than our statistical 
errors of about 10\% for this quantity. 
This is a pleasing feature for lattice calculations 
and it is also confirmed by another dynamical calculation \cite{SESAM_potential}  which could not resolve any change of the static hybrid potential.
In this picture one can also understand why relativistic corrections and 
discretisation errors are so small - in the flat potential the quarks 
move very slowly and are widely separated, hence they are not very 
sensitive to the details of the lattice cutoff.
This is in contrast to the S-state splittings, where 
dynamical effects are to be expected because of their sensitivity 
to the physics at short distance scales.
Here we observe a 5\% shift of $R_{2S}$, in line with
an earlier calculation at similar lattice spacing \cite{spitz}.
It is important to realize that this accounts for only 
part of the dynamical effects in the real world. 
As it has been argued before \cite{shigemitsu}, our sea
quark mass is sufficiently small for the purpose of bottomonium 
calculations, but it is apparent that large gluon momenta can also 
excite strange quarks from the vacuum - a full description ought to include
three dynamical flavours. 

For a short-range quantity like $R_{2S}$ it is equally important
to study scaling violations. Our previous work \cite{cppacs_nf2} 
shows that discretisation errors may also account for a fraction of 
the remaining $\le 10\%$  deviation from experiment.
We leave a more precise determination of those two different
sources of systematic errors to future studies.
For $R_{1B}$ we have convincing evidence from previous simulations
\cite{aniso_hybrid,morning_hybrid} that at $a\approx 0.1$ fm 
scaling violations are negligible for spatially large hybrid states.
In the following we take the deviation of $R_{2S}$ from its 
experimental value as a conservative estimate for residual 
systematic errors in our 2-flavour simulation.
 
\begin{figure}
\begin{center}
\leavevmode
\hbox{\epsfxsize = 10 cm \epsffile{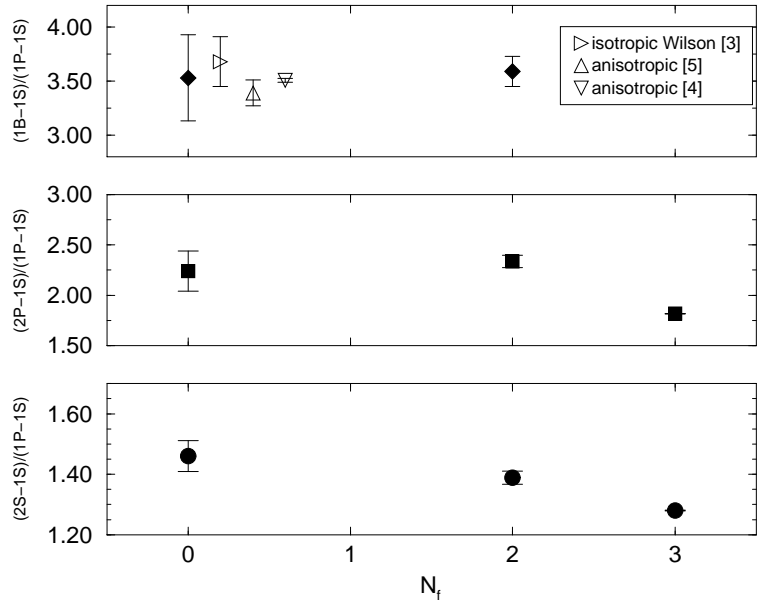}}
\end{center}
\caption{Dependence of $R_X$ on numbers of dynamical flavours, $N_f$.
The experimental values are plotted at $N_f=3$. For $R_{1B}$ we also 
show quenched results from other groups as triangles (offset for clarity).}
\label{fig:RX_vs_Nf}
\end{figure}

In Figure \ref{fig:RX_vs_Nf} we plot our results against $N_f$, along
with previous quenched estimates from different lattice spacings and 
a variety of isotropic/anisotropic lattice actions.
The stability of all these results and the good agreement with our 
value for 2-flavour full QCD provides support to the arguments above.
Converting our dynamical result into dimensionful units we quote
10.977(61)(62) GeV for the lowest lying hybrid, where the first error 
is statistical and the second denotes the systematic uncertainty 
in the determination of the lattice spacing.
 
Taking the conservative average of all points in Figure \ref{fig:RX_vs_Nf}, 
we find 11.02(18) GeV as the lattice prediction for the mass of the lowest 
spin-averaged bottomonium hybrid.
We want to stress the importance of this result, as it shows that even after
introduction of two dynamical flavours the lowest lying hybrid will be above
the $B\bar B$ threshold (10.56 GeV), where a number of experiments 
are currently running.
In fact our prediction is intriguingly close to the 
$B\bar B^{**}$ threshold ($\approx 11.0$ GeV), below which hybrid states
are thought to be very narrow.

Our results show that the error from the quenched approximation
is small within the  statistical error of isotropic lattice calculations 
($\approx 1\%$ of the bound state energy).
To study systematic errors for hybrid states at an even higher level 
of accuracy is possible on anisotropic lattices, but it remains to 
be seen how dynamical quarks can be implemented in such a framework.

The calculations were done using the supercomputing facilities
at the Center for Computational Physics at the University of Tsukuba.
This work is supported in part by the Grants-in-Aid of Ministry of
Education (No. 09304029). TM, HPS and AAK are supported by the 
JSPS Research for Future Program, and SE and KN are JSPS Research Fellows.

\begin{table}
\begin{tabular}{cccccccc}
$(\beta,\kappa)$ & $(N_s,N_t)$ & $a_\sigma$ [fm] & $a_{1P-1S}$ [fm]& $(am_Q,n)$ & $u_{0L}$ & config.   \\
\hline 
(2.10, 0.1382)  & (24,48)  & 0.11293(44) & 0.10403(80)& (2.24,2)          & 0.854604 & 400\\
(2.528,quenched)& (24,48)  & 0.11319(97) & 0.0883(16) & (1.92,2) (2.24,2) & 0.876700 & 192  \\
\end{tabular}
\caption{Simulation parameters. For the dynamical run we measured
$am_\pi/am_\rho=0.5735(48)$.}
\label{tab:parameters}
\end{table}

\begin{table}
\begin{tabular}{ccccl}
$(\beta,N_f)$   & (2.528, 0) & (2.528, 0) & (2.10, 2) & experiment \\
$am_Q$ & 1.92     & 2.24     & 2.24 & \\
\hline 
$aM_{\rm kin}$       & 4.008(16)  & 4.628(17)  & 4.642(18)  & \\
1P-1S                & 0.2057(34) & 0.2007(37) & 0.2318(16) & 439.37(13) MeV \\
$M_{\rm kin}$ [GeV]  & 8.74(15)   & 10.35(19)  & 8.805(76)  & 9.46030(26) \\
\hline
(2S-1S)/(1P-1S)  & 1.461(51)  & 1.459(48) & 1.389(22) & 1.2802(16)\\
(3S-1S)/(1P-1S)  & 3.12(27)   & 3.15(23)  & 3.255(98) & 2.0353(24) \\
(2P-1S)/(1P-1S)  & 2.28(18)   & 2.21(19)  & 2.336(60) & 1.8186(20) \\
(3P-1S)/(1P-1S)  & 4.36(40)   & 4.47(42)  & 4.94(17)  & -- \\
(1B-1S)/(1P-1S)  & 3.43(45)   & 3.63(39)  & 3.59(14)  & -- \\
(2B-1S)/(1P-1S)  & 5.8(1.1)   & 6.1(1.1)  & 6.83(52)  & --  \\
\end{tabular}
\caption{Normalised splittings with respect to $1P-1S$. The experimental values are
taken from $n ^{3}S_1$ and the spin-averaged $nP$. The dynamical
 results come from our lightest sea quark mass ($\kappa=0.1382$).}
\label{tab:results}
\end{table}

\end{document}